\documentclass{PoS}%
\usepackage{amsmath}
\usepackage{amsfonts}
\usepackage{amssymb}
\usepackage{graphicx}%
\providecommand{\U}[1]{\protect\rule{.1in}{.1in}}
\title{%
\vspace{-2cm} {%
\begin{flushright}
\begin{minipage}{5cm}
\normalsize
KEK Preprint 2014-43 \\
CHIBA-EP-210 
\end{minipage}
\end{flushright}
}
\vspace{1cm}
Magnetic monopole and confinement/deconfinement phase transition in SU(3) Yang-Mills theory \\
}

\ShortTitle{Magnetic monopole and confinement/deconfinement phase transition in SU(3) Yang-Mills theory}

\author{\speaker{Akihiro Shibata}\\
Computing Research Center, High Energy Accelerator Research Organization (KEK) \\
and Graduate University for Advanced Studies (Sokendai), Tsukuba 305-0801, Japan\\
E-mail: \email{akihiro.shibata@kek.jp}}

\author{Kei-Ichi Kondo\\
Department of Physics, Graduate School of Science, Chiba University, Chiba 263-8522, Japan\\
E-mail: \email{kondok@faculty.chiba-u.jp}}

\author{Seikou Kato\\
Fukui National College of Technology, Sabae, Fukui 916-8507, Japan\\
E-mail: \email{skato@fukui-nct.ac.jp}}

\author{Toru Shinohara\\
Department of Physics, Graduate School of Science, Chiba University, Chiba 263-8522, Japan\\
E-mail: \email{sinohara@graduate.chiba-u.jp}}

\abstract{%
We have proposed the non-Abelian dual superconductivity in SU(3) Yang-Mills  theory for the mechanism of quark confinement,
and we presented the numerical evidences in preceding lattice conferences by using the proposed gauge link decomposition 
to extract magnetic monopole in the gauge invariant way. In this talk, we focus on the dual Meissner effects in view of the magnetic monopole in SU(3) Yang-Mills theory. 
We measure the chromoelectric and chromomagnetic flux due to a pair of quark and antiquark source at finite temperature. 
Then, we measure the correlation function of Polyakov loops and Polyakov loop average  at various temperatures, 
and investigate chromomagnetic monopole current induced by chromo-magnetic flux in both confinement 
and deconfinement phase. We will discuss the role of the chromoelectric  monopole in confinement/deconfinement phase transition.
}

\FullConference{The 32nd International Symposium on Lattice Field Theory,\\
		23-28 June, 2014\\
		Columbia University New York, NY}

\begin{document}
\section{Introduction}

The dual superconductivity is a promising mechanism for quark confinement
\cite{dualSC}. To establish the dual superconductivity picture, we must show
the magnetic monopoles play the dominant role in quark confinement. We have
presented a new formulation Yang-Mills (YM) theory and proposeed the
non-Abelain dual superconductivity picture for $SU(3)$ Yang-Mills theory (for
a review see \cite{newFornulation}). We have presented a lattice version of a
new formulation of $SU(N)$ YM theory\cite{KSM05}, that gives the decomposition
of a gauge link variable $U_{x,\mu}=X_{x,\mu}V_{x,\mu}$ , which is suited for
extracting the dominant mode, $V_{x,\mu}$, for quark confinement in the gauge
independent way. In the case of the $SU(2)$ YM theory, the decomposition of
the gauge link variable is given by a compact representation of the
Cho-Duan-Ge-Faddeev-Niemi (CDGFN) decomposition \cite{CFNS-C} on a lattice
\cite{ref:NLCVsu2}\cite{ref:NLCVsu2-2}\cite{kato:lattice2009}. For the $SU(N)$
YM theory, the new formula for the decomposition of a gauge link variable is
constructed as an extension of the $SU(2)$ case. Our formulation can overcome
the problems in the Abelian projection method: the magnetic monopole dominant
is obtained only in special gauges such as the maximal Abelian (MA) gauge and
the Laplacian Abelian gauge, and the Abelian projection itself breaks the
gauge symmetry as well as color symmetry (global symmetry).

To the SU(3) YM theory, we have applied the minimal option. The minimal option
is obtained for the stability group of $\tilde{H}=U(2)\cong SU(2)\times U(1)$,
which is suitable for the Wilson loop in the fundamental representation. This
fact is derived from the non-Abelian Stokes\ theorem \cite{KondoNAST}. Then,
we have demonstrated the gauge-independent (invariant) restricted $V$-field
dominance (or conventionally called Abelian\ dominance) and the gauge
independent non-Abelian magnetic monopole dominance \cite{SCGTKKS08L}%
\cite{lattice2008}\cite{lattice2009}\cite{lattice2010}\cite{abeliandomSU(3)}.
The dual Meissner effect in YM theory must be examined by measuring the
distribution of the chromoelectric field strength (or chromo flux) as well as
the magnetic monopole current created by a static quark-antiquark pair. In the
$SU(2)$\ case, the extracted field corresponding to the stability group
$\tilde{H}=U(1)$ shows the dual Meissner effect \cite{AbelianDomSU(2)}, which
is a gauge invariant version of the Abelian projection in MA gauge. In the
$SU(3)$ case, there are many works on chromo flux by using Wilson line/loop
operator, e.g., \cite{Cardaci2011}\cite{Cardso}\cite{CeaCosmail2012}. At the
previous conference, we have demonstrated the non-Abelian dual Meissner
effect\cite{lattice2012}. By applying our new formulation to the $SU(3)$ YM
theory, we have given the numerical evidence of the non-Abelian dual Meissner
effect claimed by us, and found the chromoelectric flux tube by measuring the
chromo flux created by a static quark-antiquark pair. We have determined that
the type of vacuum for $SU(3)$ YM theory is of type I, which is in sharp
contrast to the $SU(2)$ case: the border of type I and type II
\cite{DMeisner-TypeI2013} or of week type I \cite{AbelianDomSU(2)}.

In this talk, we focus on the confinement/deconfinement phase transition and
the non-Abelian dual superconductivity at finite temperature: We measure a
Polyakov loop average and correlation functions of the Polyakov loops which
are defined for both the original YM\ field and extracted $V$-field to examine
the $V$-field dominance in the Polyakov loop at finite temperature. Then, we
measure the chromoelectric flux between a pair of static quark and antiquark
of the Polyakov loops, and investigate its relevance to the phase transition
and the non-Abelian dual Meissner effect.

\section{Method}

We introduce a new formulation of the lattice YM theory in the minimal option,
which extracts the dominant mode of the quark confinement for $SU(3)$ YM
theory\cite{abeliandomSU(3),lattice2010}, since we consider the quark
confinement in the fundamental representation. Let $U_{x,\mu}=X_{x,\mu
}V_{x,\mu}$ be a decomposition of the YM link variable $U_{x,\mu}$, where
$V_{x,\mu}$ could be the dominant mode for quark confinement, and $X_{x,\mu}$
the remainder part. The YM field and the decomposed new variables are
transformed by full $SU(3)$ gauge transformation $\Omega_{x}$ such that
$V_{x,\mu}$ is transformed as the gauge link variable and $X_{x,\mu}$ as the
site variable:
\begin{subequations}
\label{eq:gaugeTransf}%
\begin{align}
U_{x,\mu}  &  \longrightarrow U_{x,\nu}^{\prime}=\Omega_{x}U_{x,\mu}%
\Omega_{x+\mu}^{\dag},\\
V_{x,\mu}  &  \longrightarrow V_{x,\nu}^{\prime}=\Omega_{x}V_{x,\mu}%
\Omega_{x+\mu}^{\dag},\text{ \ }X_{x,\mu}\longrightarrow X_{x,\nu}^{\prime
}=\Omega_{x}X_{x,\mu}\Omega_{x}^{\dag}.
\end{align}
The decomposition is given by solving the defining equation:
\end{subequations}
\begin{subequations}
\label{eq:DefEq}%
\begin{align}
&  D_{\mu}^{\epsilon}[V]\mathbf{h}_{x}:=\frac{1}{\epsilon}\left[  V_{x,\mu
}\mathbf{h}_{x+\mu}-\mathbf{h}_{x}V_{x,\mu}\right]  =0,\label{eq:def1}\\
&  g_{x}:=e^{i2\pi q/3}\exp(-ia_{x}^{0}\mathbf{h}_{x}-i\sum\nolimits_{j=1}%
^{3}a_{x}^{(j)}\mathbf{u}_{x}^{(j)})=1, \label{eq:def2}%
\end{align}
where $\mathbf{h}_{x}$ is an introduced color field $\mathbf{h}_{x}%
=\xi(\lambda^{8}/2)\xi^{\dag}$ $\in\lbrack SU(3)/U(2)]$ with $\lambda^{8}$
being the Gell-Mann matrix and $\xi$ an $SU(3)$ group element. The variable
$g_{x}$ is an undetermined parameter from Eq.(\ref{eq:def1}), $\mathbf{u}%
_{x}^{(j)}$ 's are $su(2)$-Lie algebra valued, and has $q_{x}$ an integer
value $\ 0,1,2$. These defining equations can be solved exactly
\cite{exactdecomp}, and the solution is given by
\end{subequations}
\begin{subequations}
\label{eq:decomp}%
\begin{align}
X_{x,\mu}  &  =\widehat{L}_{x,\mu}^{\dag}\det(\widehat{L}_{x,\mu})^{1/3}%
g_{x}^{-1},\text{ \ \ \ }V_{x,\mu}=X_{x,\mu}^{\dag}U_{x,\mu}=g_{x}\widehat
{L}_{x,\mu}U_{x,\mu},\\
\widehat{L}_{x,\mu}  &  =\left(  L_{x,\mu}L_{x,\mu}^{\dag}\right)
^{-1/2}L_{x,\mu},\text{ \ \ \ \ \ \ }L_{x,\mu}=\frac{5}{3}\mathbf{1}+\frac
{2}{\sqrt{3}}(\mathbf{h}_{x}+U_{x,\mu}\mathbf{h}_{x+\mu}U_{x,\mu}^{\dag
})+8\mathbf{h}_{x}U_{x,\mu}\mathbf{h}_{x+\mu}U_{x,\mu}^{\dag}\text{ .}%
\end{align}
Note that the above defining equations correspond to the continuum version:
$D_{\mu}[\mathcal{V}]\mathbf{h}(x)=0$ and $\mathrm{tr}(\mathbf{h}%
(x)\mathcal{X}_{\mu}(x))$ $=0$, respectively. In the naive continuum limit, we
have reproduced the decomposition $\mathbf{A}_{\mathbf{\mu}}(x)=\mathbf{V}%
_{\mu}(x)+\mathbf{X}_{\mu}(x)$ in the continuum theory \cite{exactdecomp} as
\end{subequations}
\begin{subequations}
\begin{align}
\mathbf{V}_{\mu}(x)  &  =\mathbf{A}_{\mathbf{\mu}}(x)-\frac{4}{3}\left[
\mathbf{h}(x),\left[  \mathbf{h}(x),\mathbf{A}_{\mathbf{\mu}}(x)\right]
\right]  -ig^{-1}\frac{4}{3}\left[  \partial_{\mu}\mathbf{h}(x),\mathbf{h}%
(x)\right]  ,\\
\mathbf{X}_{\mu}(x)  &  =\frac{4}{3}\left[  \mathbf{h}(x),\left[
\mathbf{h}(x),\mathbf{A}_{\mathbf{\mu}}(x)\right]  \right]  +ig^{-1}\frac
{4}{3}\left[  \partial_{\mu}\mathbf{h}(x),\mathbf{h}(x)\right]  .
\end{align}

The decomposition is uniquely obtained as the solution (\ref{eq:decomp}) of
Eqs.(\ref{eq:DefEq}), if color fields$\{\mathbf{h}_{x}\}$ are obtained. To
determine the configuration of color fields, we use the reduction condition to
formulate the new theory written by new variables ($X_{x,\mu}$,$V_{x,\mu}$)
which is equipollent to the original YM theory. Here, we use the reduction
functional:
\end{subequations}
\begin{equation}
F_{\text{red}}[\mathbf{h}_{x}]=\sum_{x,\mu}\mathrm{tr}\left\{  (D_{\mu
}^{\epsilon}[U_{x,\mu}]\mathbf{h}_{x})^{\dag}(D_{\mu}^{\epsilon}[U_{x,\mu
}]\mathbf{h}_{x})\right\}  , \label{eq:reduction}%
\end{equation}
and then color fields $\left\{  \mathbf{h}_{x}\right\}  $ are obtained by
minimizing the functional (\ref{eq:reduction}).

\section{Lattice result}

\begin{figure}[ptb]
\begin{center}
\includegraphics[
height=5.5cm, angle=270]
{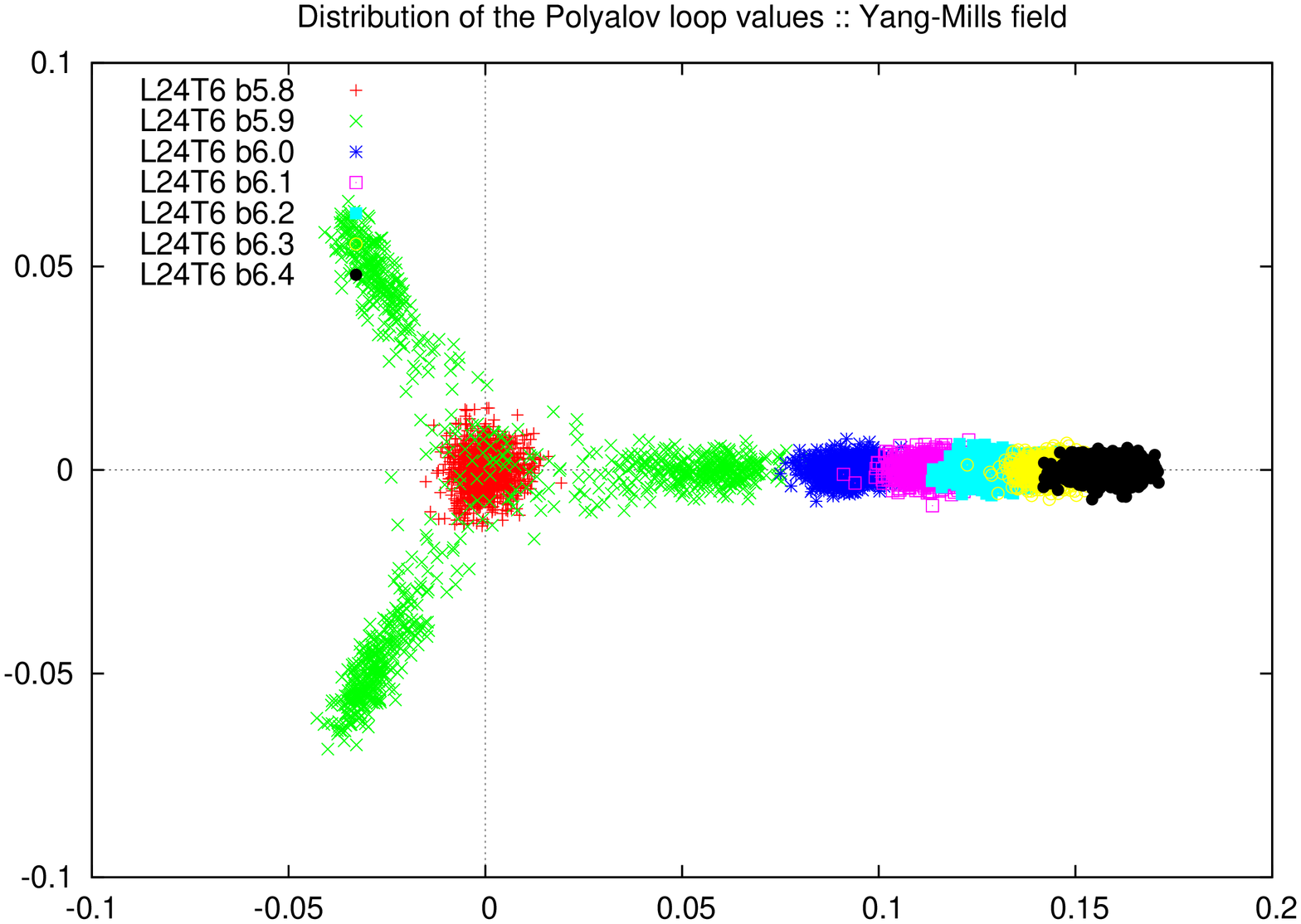} \ \ \includegraphics[
height=5.5cm, angle=270]
{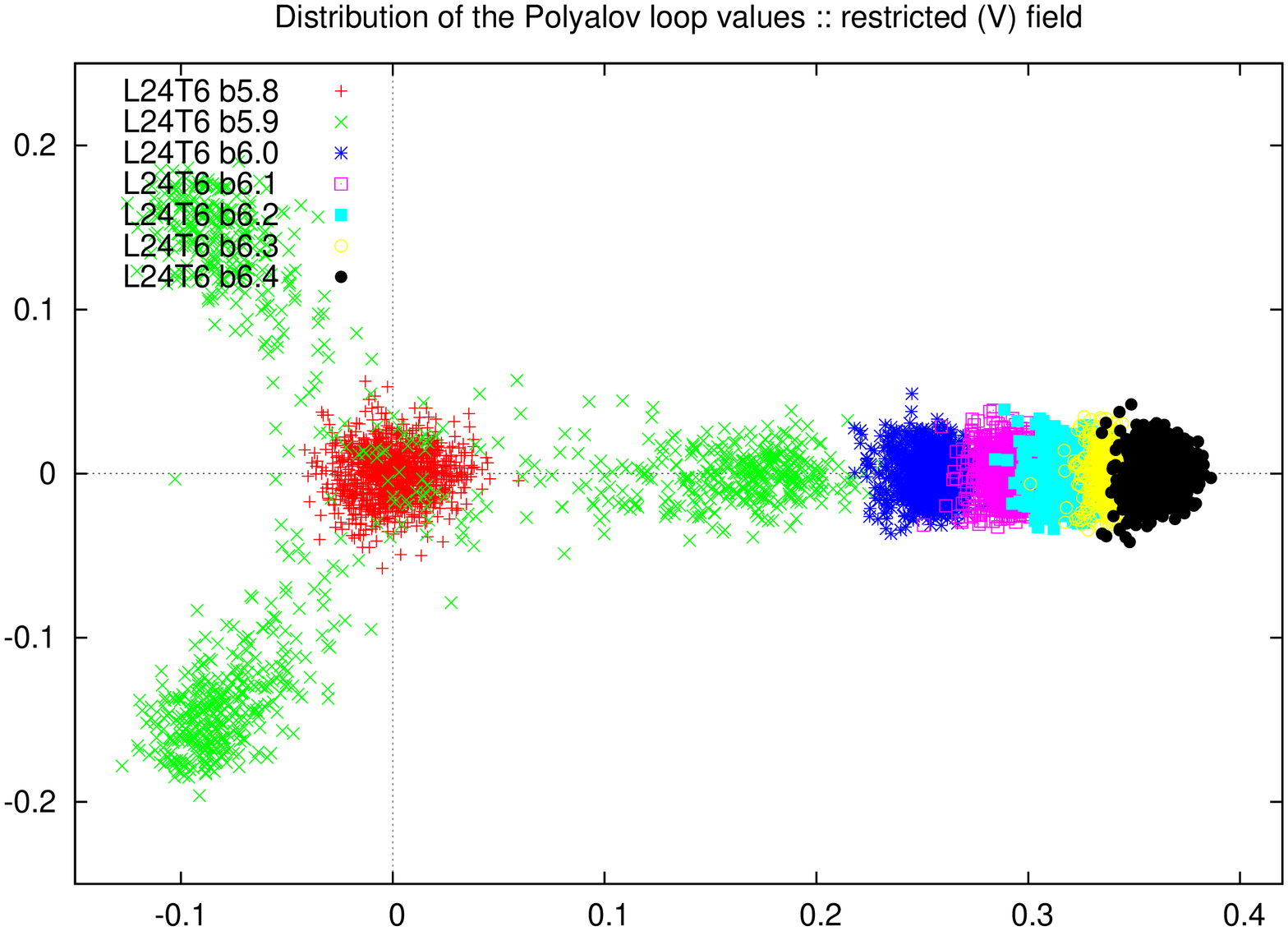}
\end{center}
\caption{{}The distribution of the space-averaged Polyakov loop for each
configuration:(Left) For the YM field. (Right) For the restricted field. }%
\label{Fig:PLP}%
\end{figure}

\begin{figure}[ptb]
\begin{center}
\includegraphics[
height=6cm, angle=270]
{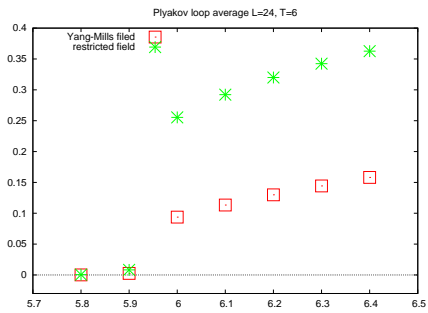}
\end{center}
\caption{{}Parameter $\beta$ dependence of the Polyakov loop average: Red
plots show $\left\langle P_{U}\right\rangle $ v.s $\beta$, green ones
$\left\langle P_{V}\right\rangle $ v.s $\beta.$}%
\label{fig:PLaverage}%
\end{figure}We generate YM gauge configurations$\{U_{x,\mu}\}$ at finite
temperature using the standard Wilson action. We set up a lattice $L^{3}\times
N_{T}$ ($L=24,N_{T}=6$) and the control the temperature by changing the
parameter $\beta$: $\beta=5.8$, $5.9$, $6.0$, $6.1$, $6.2$, $6.3.$ We generate
500 configurations for each $\beta$. In the measurement of the Polyakov loop
and Wilson loop, we apply the APE smearing technique to reduce noises
\cite{APEsmear}. The gauge link decomposition $U_{x,\mu}=X_{x,\mu}V_{x,\mu}$
is obtained by the formula (\ref{eq:decomp}) given in the previous section,
after the color field configuration $\{\mathbf{h}_{x}\}$ is obtained by
solving the reduction condition of minimizing the functional
eq(\ref{eq:reduction}) for each gauge configuration$\{U_{x,\mu}\}$.

Figure \ref{Fig:PLP} shows the distribution of space-averaged Polyakov loops
for each configuration:
\begin{equation}
P_{U}:=L^{-3}\sum\nolimits_{\{\vec{x}\}}\mathrm{tr}\left(  \prod
\nolimits_{t=1}^{N_{T}}U_{(\vec{x},t),4}\right)  ,\text{ \ \ \ }P_{V}%
:=L^{-3}\sum\nolimits_{\{\vec{x}\}}\mathrm{tr}\left(  \prod\nolimits_{t=1}%
^{N_{T}}V_{(\vec{x},t),4}\right)  . \label{eq:PLP}%
\end{equation}
The left panel Fig.\ref{Fig:PLP} shows the distribution of $P_{U}$ for the
YM\ field for each configuration, and the right panel shows the distribution
of $P_{V}$ for the restricted field ($V$-field) for the relevant
configurations. Then, we obtain the Polyakov loop average, which is the
conventional order parameter for confinement and deconfinement phase
transition in $SU(3)$ YM\ theory. Figure \ref{fig:PLaverage} shows combined
polts of the Polyakov loop average for the YM field $\left\langle
P_{U}\right\rangle $ and the restricted field $\left\langle P_{V}\right\rangle
$. Each plot shows the same critical temperature of confinement/deconfinement
phase transition. These show the extracted $V$-field reproduces the phase
transition at finite temperature. We can also show the restricted field
($V$-field) dominance in the Polyakov loop correlation functions, which was
presented at the last conference\cite{lattice2013}.

\begin{figure}[ptb]
\begin{center}
\includegraphics[
height=4.5cm]
{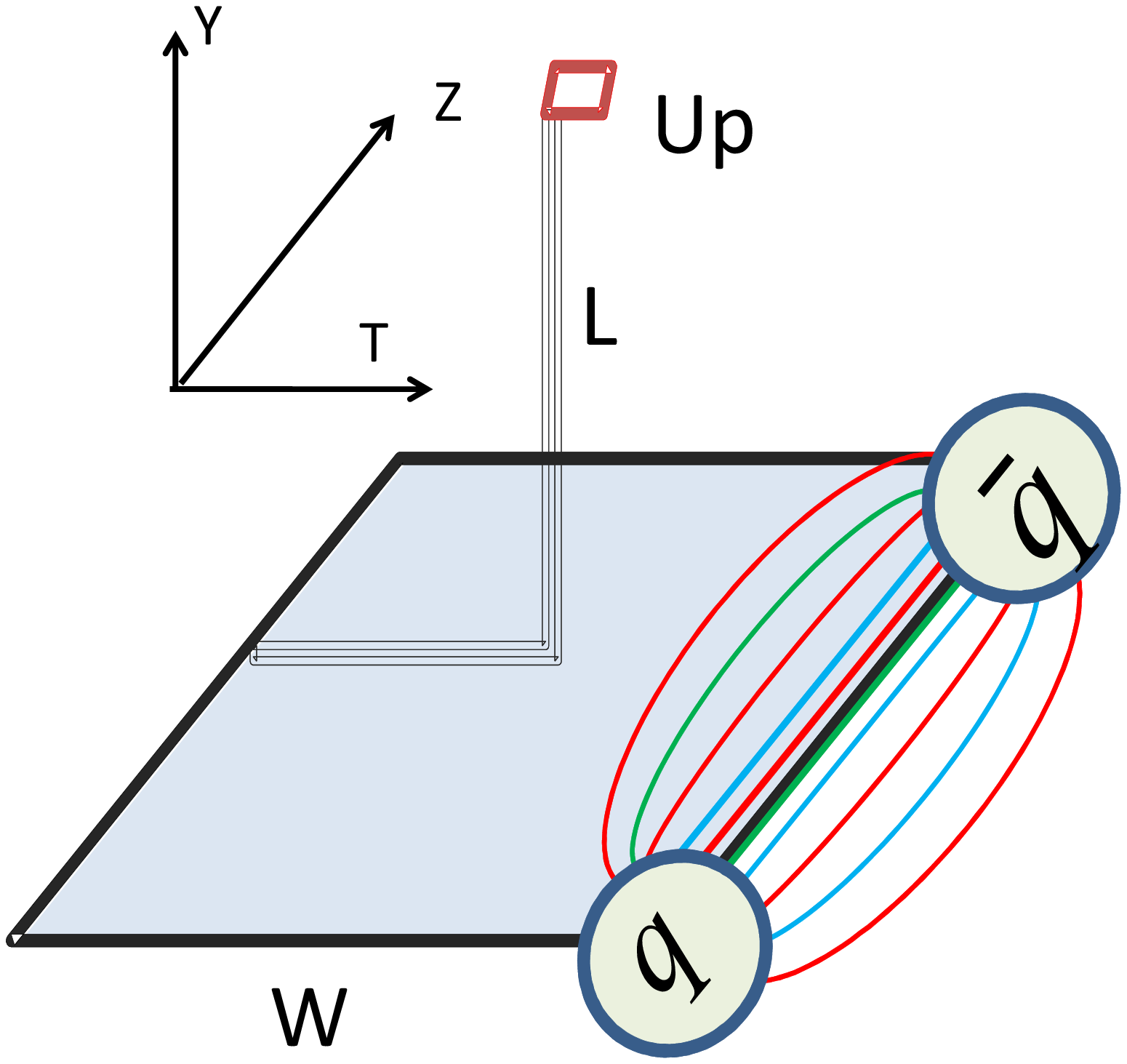} \ \ \includegraphics[
height=3.5cm]
{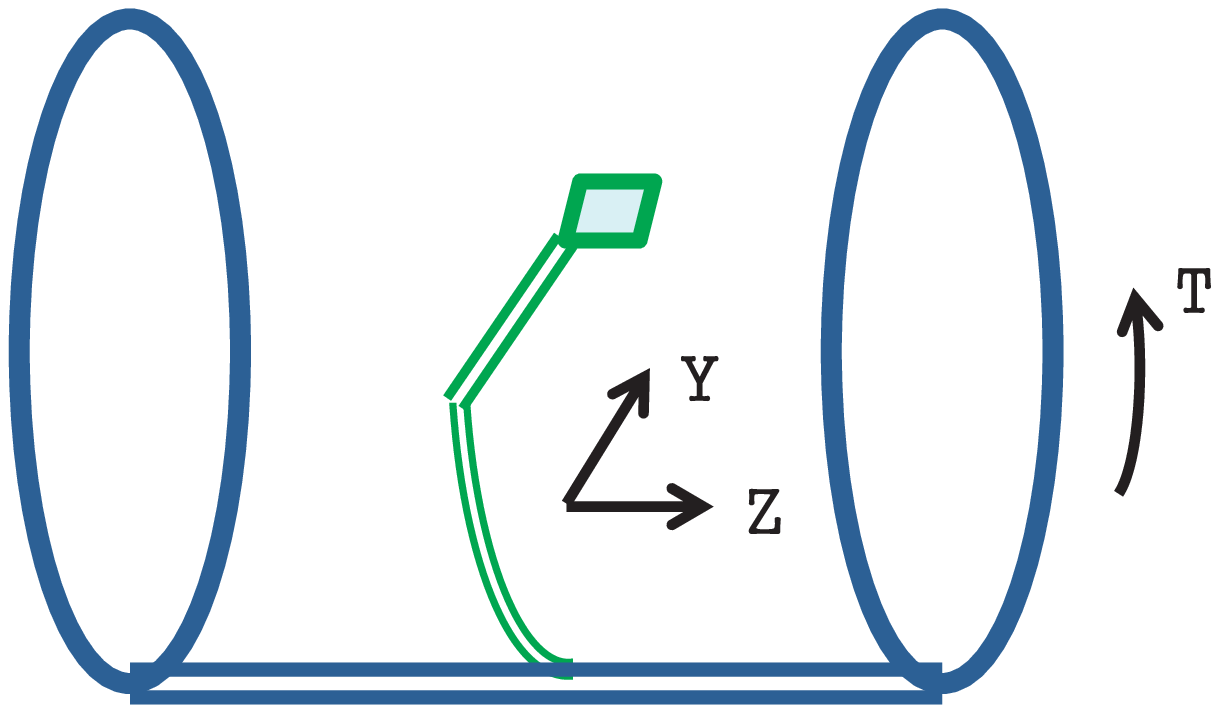}
\end{center}
\caption{(Left) The connected operator $WLU_{p}L^{\dag}$ between a plaquette
$U_{p}$ and the Wilson loop $W.$ (Right) Measurement of the chromo-flux at
finite temperature via Polyakov loop. }%
\label{fig:measure1}%
\end{figure}

Next, we investigate the non-Abelian dual Meissner effect at finite
temperature. Note that at finite temperature we must use the operator with the
same size in the temporal direction, and the quark and antiquark pair is
replaced by a pair of the Polyakov loop with the opposite direction. To
investigate the chromo flux, we use the gauge invariant correlation function
which is used at zero temperature. The chromo flux created by a
quark-antiquark pair is measured by using a gauge-invariant connected
correlator of the Wilson loop \cite{Giacomo}:%
\begin{equation}
\rho_{W}:=\frac{\left\langle \mathrm{tr}\left(  U_{p}L^{\dag}WL\right)
\right\rangle }{\left\langle \mathrm{tr}\left(  W\right)  \right\rangle
}-\frac{1}{3}\frac{\left\langle \mathrm{tr}\left(  U_{p}\right)
\mathrm{tr}\left(  W\right)  \right\rangle }{\left\langle \mathrm{tr}\left(
W\right)  \right\rangle }, \label{eq:Op}%
\end{equation}
where $W$ represents the source of a quark-antiquark pair settled by the
Wilson loop in Z-T plane, $U_{p}$ a plaquette variable as the probe operator
for measuring the field strength, and $L$ the Wilson line connecting the
source $W$ and the probe $U_{p}.$ (see the left panel\ of Figure
\ref{fig:measure1}). The symbol $\left\langle \mathcal{O}\right\rangle $
denotes the average of the operator $\mathcal{O}$ over the space and the
ensemble of the configurations. Note that this is sensitive to the field
strength rather than the disconnected one. Indeed, in the naive continuum
limit, the connected correlator $\rho_{W}$ is given by $\ \rho_{W}%
\overset{\varepsilon\rightarrow0}{\simeq}g\epsilon^{2}\left\langle
\mathcal{F}_{\mu\nu}\right\rangle _{q\bar{q}}:=\frac{\left\langle
\mathrm{tr}\left(  g\epsilon^{2}\mathcal{F}_{\mu\nu}L^{\dag}WL\right)
\right\rangle }{\left\langle \mathrm{tr}\left(  W\right)  \right\rangle
}+O(\epsilon^{4})$. Thus, the chromo field strength is given by $\ F_{\mu\nu
}=\sqrt{\frac{\beta}{6}}\rho_{W}$.

\begin{figure}[ptb]
\begin{center}
\includegraphics[
height=5cm, angle=270]
{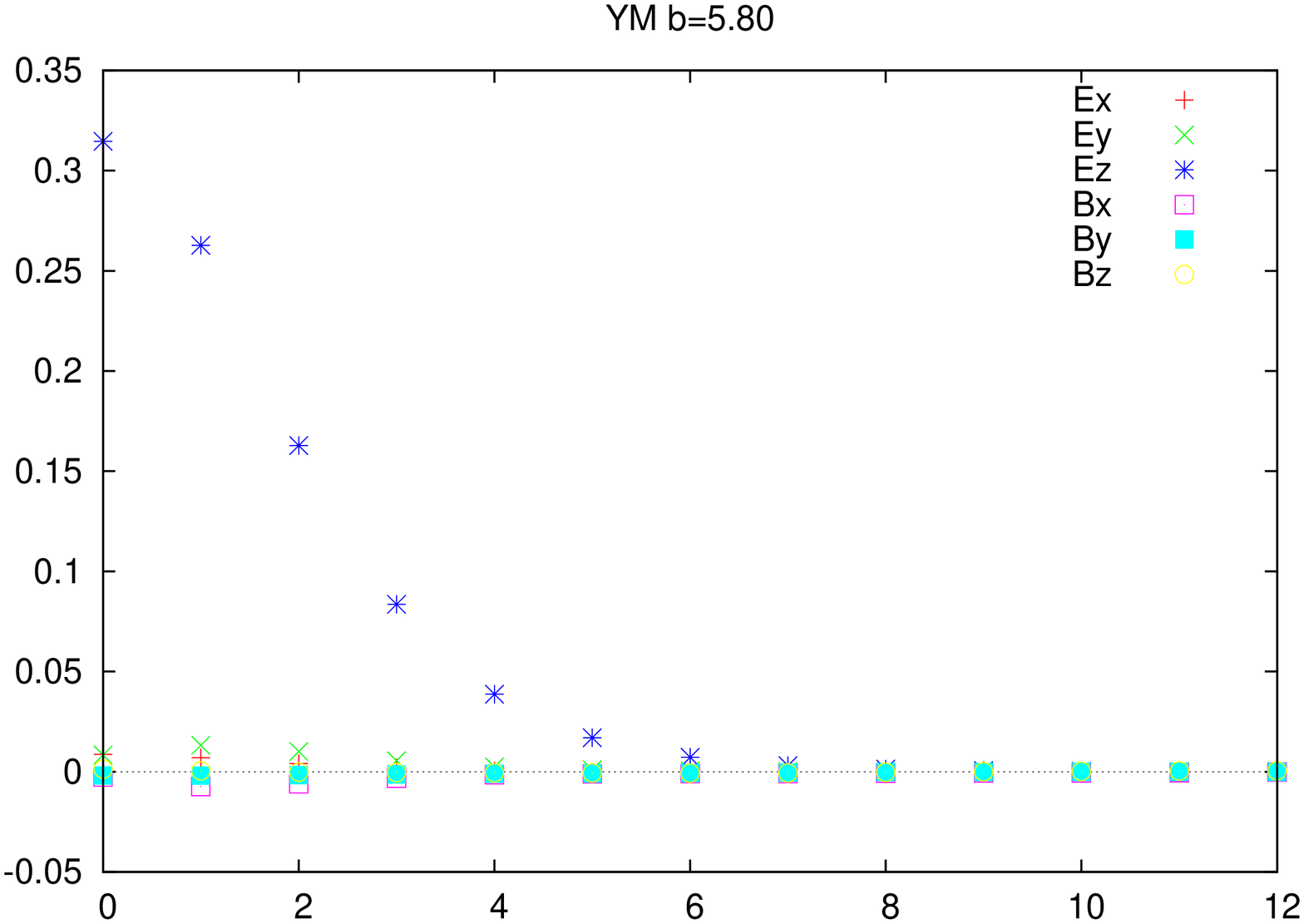}\includegraphics[
height=5cm, angle=270]
{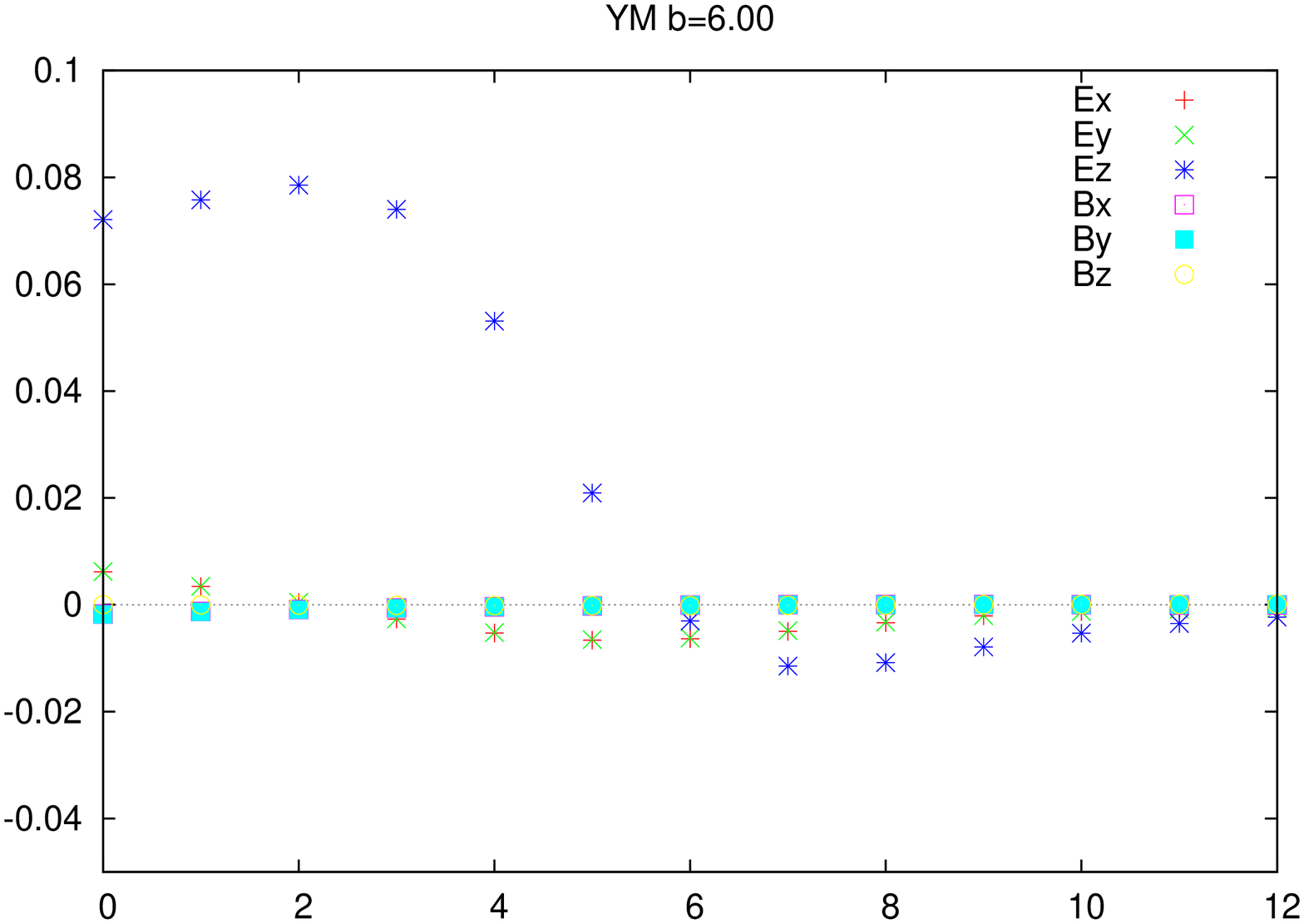}\includegraphics[
height=5cm, angle=270]
{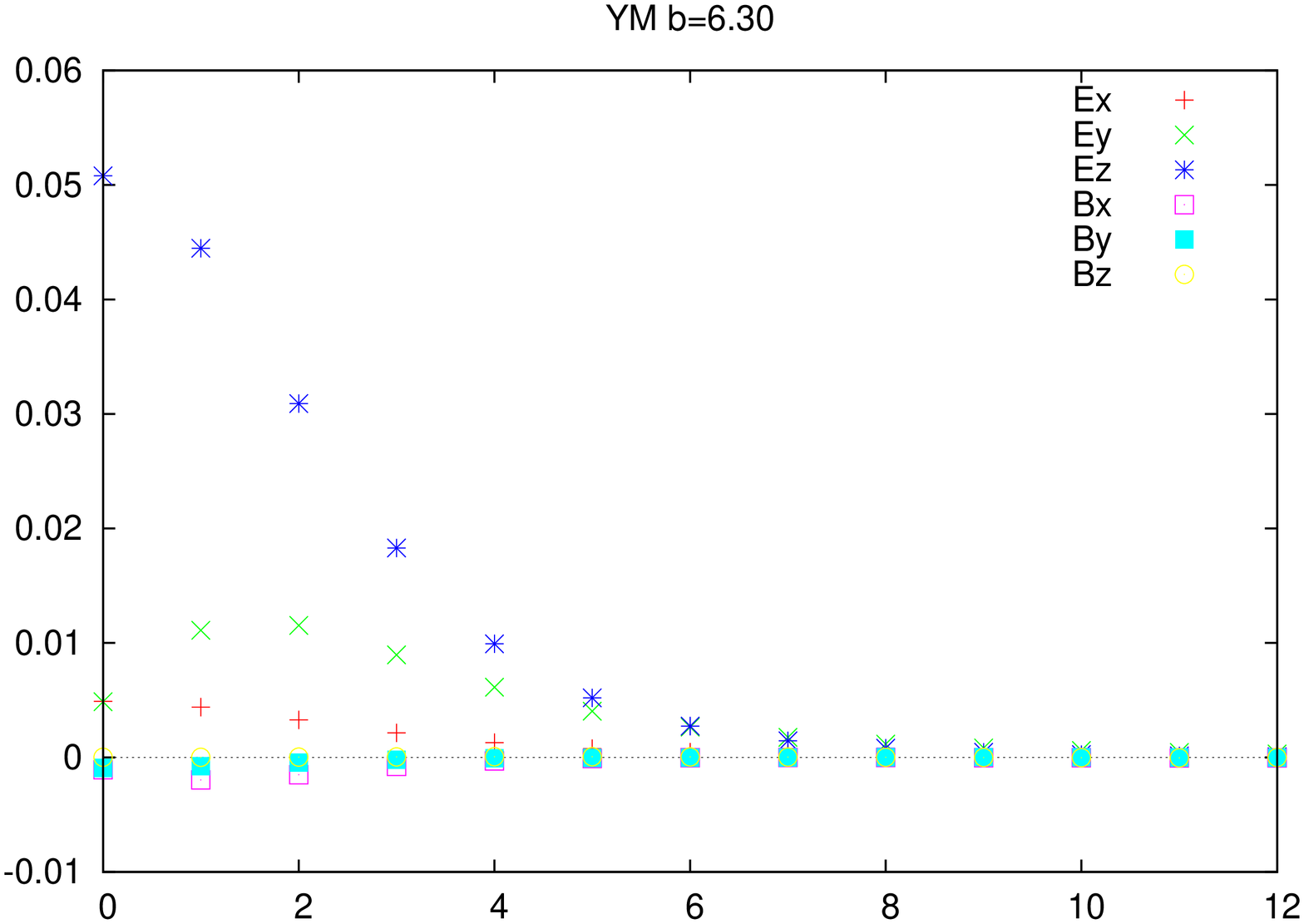}
\end{center}
\caption{{}chromo-flux created by a pair of Polyakov loops of the YM field.
(left) $\beta=5.80$, (middle )$\beta=6.00$, (right) $\beta=6.30$ }%
\label{fig:flux-YM}%
\end{figure}

\begin{figure}[ptb]
\begin{center}
\includegraphics[
height=5cm, angle=270]
{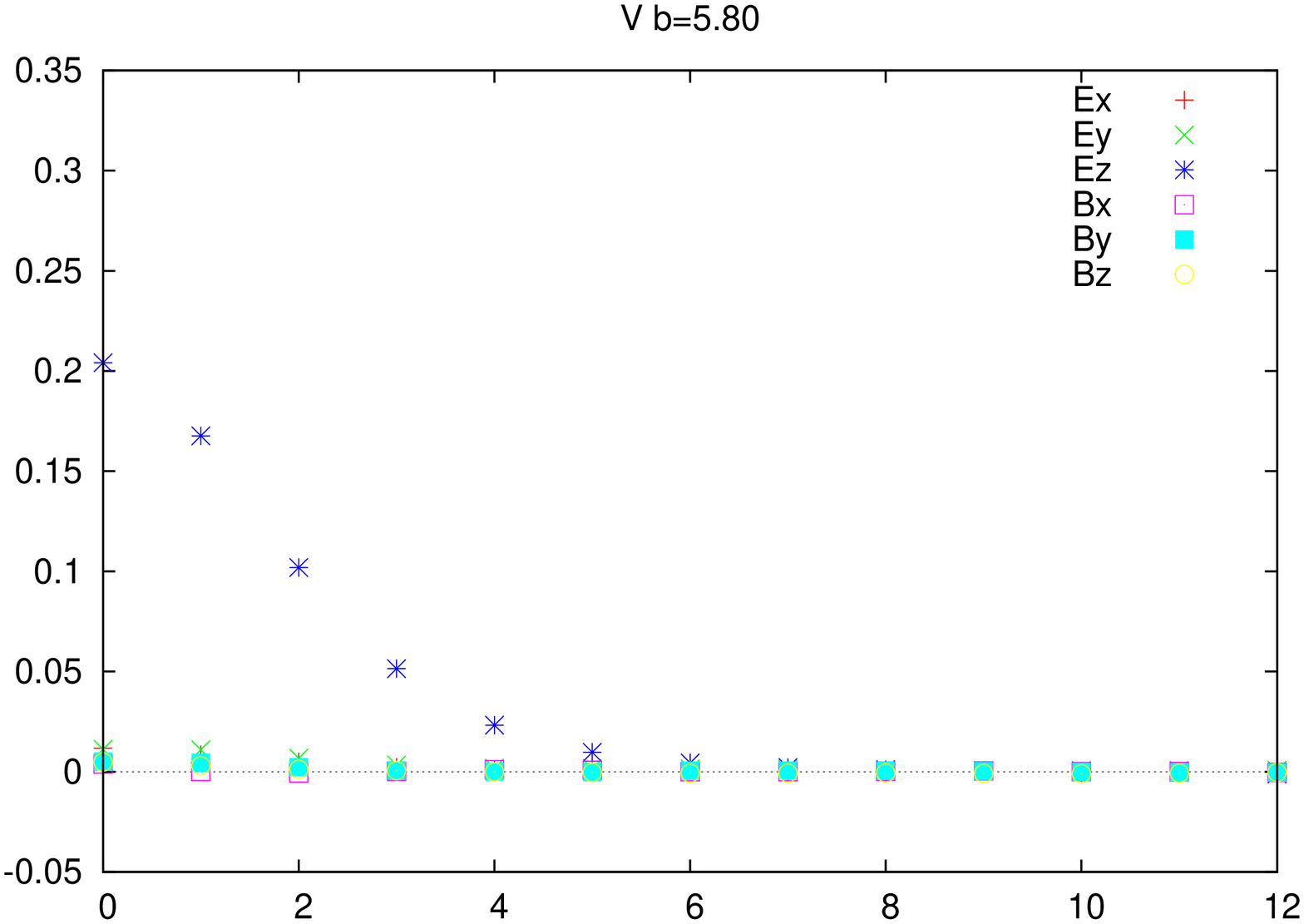}\includegraphics[
height=5cm, angle=270]
{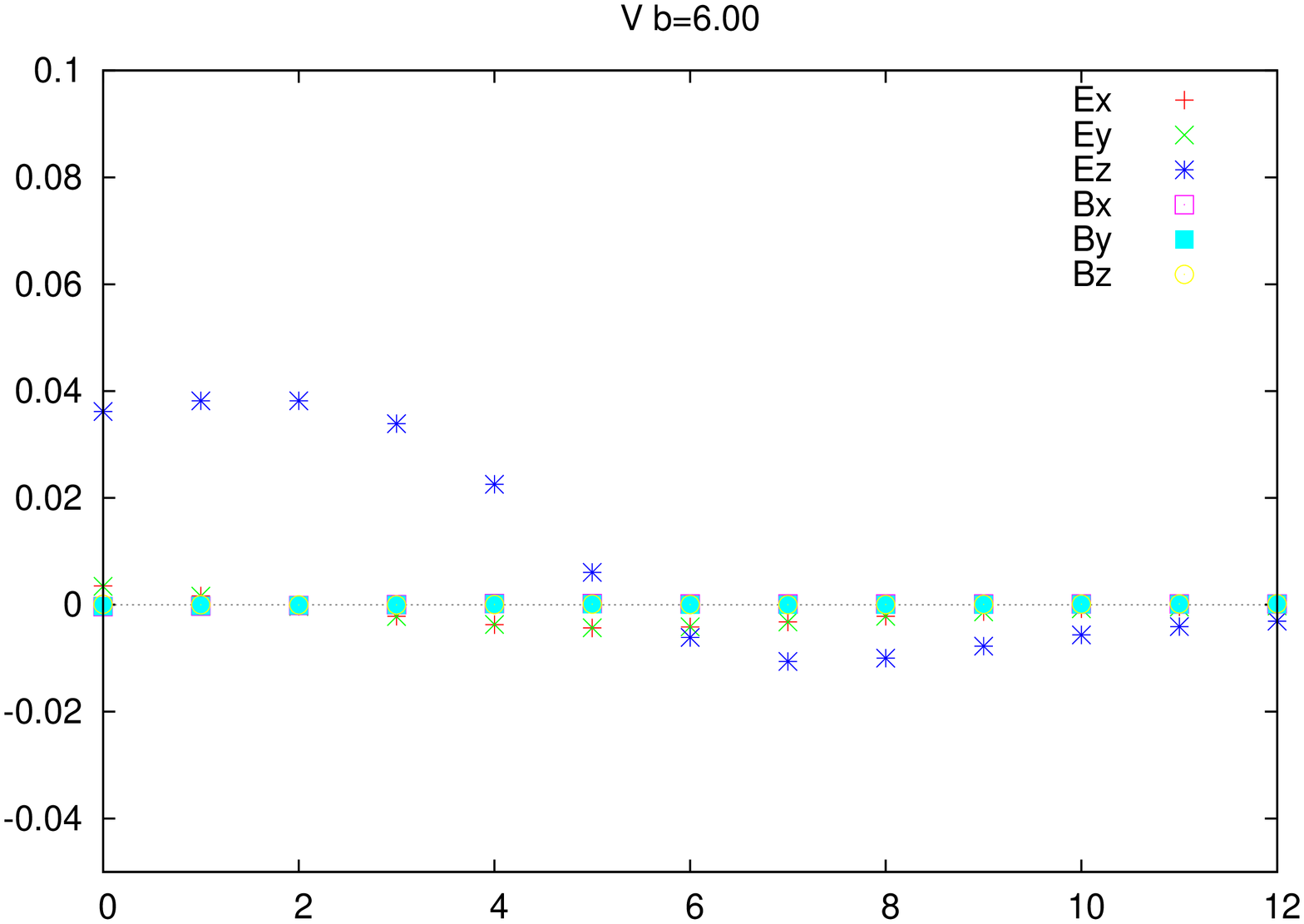}\includegraphics[
height=5cm, angle=270]
{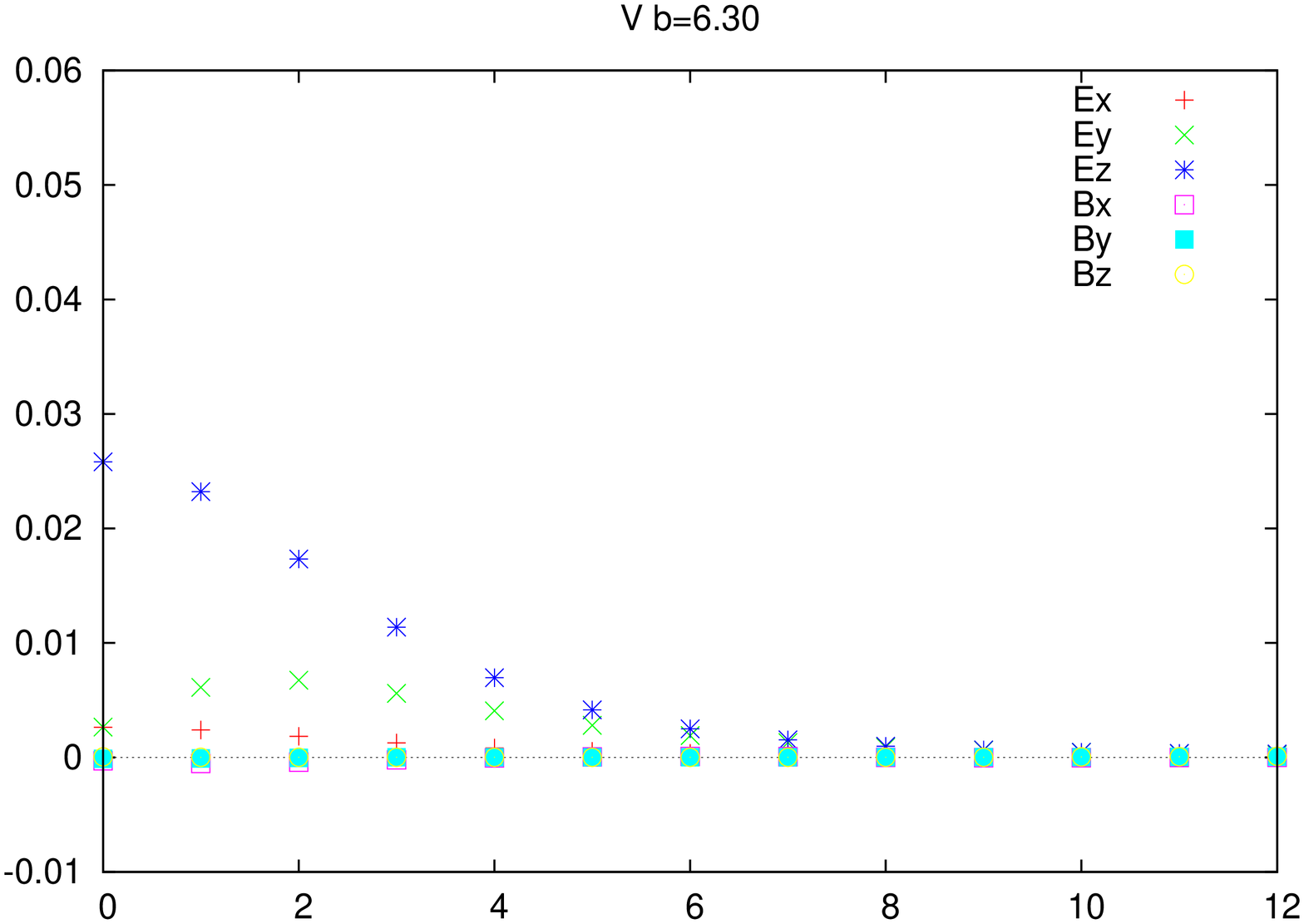}
\end{center}
\caption{{}chromo-flux created by a pair of Polyakov loops of the restricted
field. (left) $\beta=5.80$, (middle )$\beta=6.00$, (right) $\beta=6.30$ }%
\label{fig:flux-V0}%
\end{figure}

Figure \ref{fig:flux-YM} and \ref{fig:flux-V0} show the measurement of chromo
flux for Yang-Mills field and restricted field ($V$-field) at finite
temperature. The chromo-flux of quark-antiquark pair is measured on the plane
at $z=1/3R$ for a given quark at $z=0$ and an antiquark at $z=R$ by moving the
probe, $U_{p}$ or $V_{p}$ along the y-direction. We find the restricted field
dominance for the chromo-flux tube at finite temperature as well as zero temperature

At low temperature (see left panels of Fig.\ref{fig:flux-YM} and
Fig.\ref{fig:flux-V0}), we observe the chromoelectric flux tube such that only
the $E_{z}$ component in the direction connecting a quark and antiquark pair
is observed, while the other components take vanishing values. This is
consistent with the result Ref.\cite{P.Cea-L.Lenardo2014}, though they use the
different operator for the flux measurements.

At high temperature ($T>T_{C}$) (see right panels of Fig.\ref{fig:flux-YM} and
Fig.\ref{fig:flux-V0}), we can observe no more squeezing of the chromoelectric
flux tube, and non-vanishing $E_{y}$ component \ in the chromoelectric flux.
This shows the disappearance of the dual Meissner effect at high temperature.

\begin{figure}[ptbh]
\begin{center}
\includegraphics[
height=6cm, angle=270]
{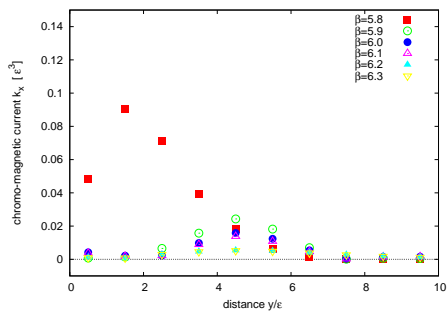}
\end{center}
\caption{The magnetic carrent (monopole) induced by \ a pair of Polyakov
loops. }%
\label{fig:m_current}%
\end{figure}

Then, we investigate the magnetic (monopole) current due to the magnetic
condensation:
\begin{equation}
k_{\mu}(x)=\frac{1}{2}\epsilon_{\mu\nu\alpha\beta}\left(  F[V]_{\alpha\beta
}(x+\hat{\nu})-F[V]_{\alpha\beta}(x)\right)  \label{eq:m_current}%
\end{equation}
Note that the magnetic monopole current eq(\ref{eq:m_current}) must have
vanishing value if there exists no magnetic monopole condensation, since the
right-hand side of eq(\ref{eq:m_current}) is the Bianchi identity. Therefore,
the magnetic monopole current can be the order parameter of dual
superconductivity. Figure \ref{fig:m_current} shows the measurements for the
magnetic current eq(\ref{eq:m_current}). We observe the appearance and
disappearance of the magnetic monopole current at low and high temperature,
respectively. 

\section{Summary and outlook}

We have investigated the non-Abelian dual Meissner effect at finite
temperature by measuring the chromo flux due to a pair of quark and antiquark
source represented by a pair of the Polyakov loops. Using our proposal for a
new formulation of Yang-Mills theory on a lattice, we ware able to extract the
dominant mode for quark confinement as the restricted field ($V$-field), and
confirmed that the restricted field dominance at finite temperature. We have
observed no more squeezing of the chromoelectric flux tube due to the dual
Meissner effect. We have also measured the magnetic (monopole) current in both
the confinement and deconfinement phase, and observed that the
confinement/deconfinement phase transition is associated to appearance and
disappearance of magnetic (monopole) current. This is the evidence that the
confinement/deconfinement phase transition is caused by
appearance/disappearance of the non-Abelian dual superconductivity.

\subsection*{Acknowledgement}

This work is supported by Grant-in-Aid for Scientific Research (C) 24540252
from Japan Society for the Promotion Science (JSPS), and also in part by JSPS
Grant-in-Aid for Scientific Research (S) 22224003. The numerical calculations
are supported by the Large Scale Simulation Program No.12-13 (FY2012), No.
12/13-20 (FY2012/13) and No.13/14-23 (FY2013-2014) of High Energy Accelerator
Research Organization (KEK).

\end{document}